\documentclass[prl,twocolumn,showpacs,preprintnumbers,aps,nofootinbib]{revtex4-1}

\usepackage{graphicx}  %commented out by MK
\usepackage{dcolumn}
\usepackage{bm}
\usepackage{amsmath,amssymb,amsfonts}
\usepackage{latexsym}
\usepackage{color}
\usepackage[colorlinks=true,urlcolor=black,anchorcolor=blue
,citecolor=blue,filecolor=blue,linkcolor=black,menucolor=blue
,linktocpage=true,pdfproducer=medialab,pdfa=true]{hyperref}
\def\nn    {\nonumber}

\begin{document}
\title{\boldmath
Searching for resonant flavor-changing charged Higgs production at the LHC}
\author{Wei-Shu Hou and Mohamed Krab}
\affiliation{Department of Physics, National Taiwan University, Taipei 10617, Taiwan}
	\bigskip
\begin{abstract}
We suggest a resonant $c \bar b \to H^+$ production search, followed by bosonic $H^+ \to W^+ H$ weak decay at the Large Hadron Collider (LHC). In the general two-Higgs-doublet model 
(G2HDM) that has flavor-changing neutral Higgs couplings, $H^+$ is resonantly produced via the top-charm $\rho_{tc} V_{tb}$ coupling at tree level, while $H^+ \to W^+ H$ weak decay occurs within the exotic second doublet, leading eventually to same-sign dilepton signals. We perform a signal-to-background analysis at the 14~TeV LHC and show that discovery seems possible with LHC Run~2 data already at hand.
\end{abstract}
	\maketitle
%%%%%%%%%%%%%%%%%%%%%%%%%%%%%%%%%%%%%%%%
%%%%%%%%%%%%%%%%%%%%%%%%%%%%%%%%%%%%%%%%%%%%%%%%
{\it Introduction.---}%%%%%%%%	
The discovery of the $125$ GeV state compatible with the Standard Model (SM)-like Higgs boson at the LHC~\cite{LHC:2012}, which belongs to a weak scalar doublet, is the most significant breakthrough in particle physics. Only the SM Higgs boson $h$ is found so far, with no sign of extra scalars. However, given the repetition of fermion weak doublets, it is {\it imperative} to pursue the existence of a second exotic scalar doublet---in particular its associated charged $H^+$ boson.

We study the $H^+$ boson in the general two-Higgs-doublet model 
(G2HDM) where the usual $Z_2$ symmetry is absent, as imposing~\cite{Glashow:1976nt} such a symmetry would be plainly {\it ad hoc}. With two identical Higgs doublets, there are three neutral scalars, the {\it CP}-even $h, H$; the {\it CP}-odd $A$; and a charged $H^\pm$ pair.

\begin{figure}[b]
	\centering
	\includegraphics[width=.25\textwidth]{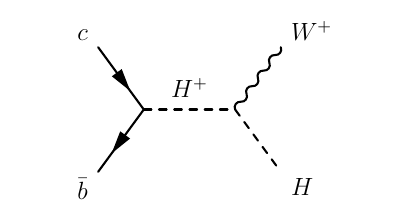}\hskip-0.1cm
	\caption{\footnotesize Diagram for $c\bar b \to H^+ \to W^+ H$, with $H \to t\bar c$.}
	\label{diag}
\end{figure}

The discovery potential of $H^+$ at the LHC has been studied in the G2HDM~\cite{Ghosh:2019exx,Hou:2024bzh}. For $m_{H^+} \sim$ 300--500 GeV, $cg \to bH^+$ is the most promising.
Compared with 2HDM-II, which arises with supersymmetry, the process is not~\cite{Ghosh:2019exx} Cabibbo-Kobayashi-Maskawa (CKM) suppressed. In this paper, we promote $s$-channel resonant $H^+$ production, the $c\bar b \to H^+$ process of Fig.~\ref{diag}, which has the same $\bar c bH^+$ coupling of $\rho_{tc}V_{tb}$, where $\rho_{tc}$ is the flavor-changing top-charm coupling and $V_{tb}$ is a CKM matrix element. We turn to the subsequent decay shortly.

Extra top Yukawa couplings $\rho_{tc}$ and $\rho_{tt}$ are not severely constrained and could be $\mathcal{O}(1)$, and they can {\it each} drive~\cite{Fuyuto:2017ewj} electroweak baryogenesis (EWBG), and hence account for the baryon asymmetry of the Universe (BAU), a big motivator. At the same time, one can evade the stringent electron electric dipole moment (eEDM) bounds of ACME~\cite{ACME:2018yjb} and JILA~\cite{Roussy:2022cmp}, by quite {\it a natural} flavor cancellation mechanism~\cite{Fuyuto:2019svr}. In addition, $\rho_{tc}$ and $\rho_{tt}$ at $\mathcal{O}(1)$ facilitate interesting production and decay processes, such as $cg \rightarrow bH^+ \to bt\bar b$~\cite{Ghosh:2019exx}, and $\bar bg \to \bar c H^+ \to \bar c t\bar b$~\cite{Hou:2024bzh}, yielding final states with at least three jets and lepton plus missing energy, giving better signal sensitivity. 
However, even if kinematically allowed, the bosonic $H^+ \to W^+ H$ decay would be negligible compared to fermionic $H^+ \to c\bar b$ and $H^+ \to t\bar b$ modes when $\rho_{tc}$ and $\rho_{tt}$ are {sizable}. 

We point out that the $H^+ \to W^+ H$ weak decay might become important {\it if both $\rho_{tc}$ and $\rho_{tt}$ are small enough}, or {\it if there is large $m_{H^+}$-$m_H$ mass splitting}, as we shall elaborate. We note that ${\rm Im}\,\rho_{tt} \simeq 0.1$ is still quite robust in driving EWBG~\cite{Fuyuto:2017ewj}.
The $s$-channel {$c\bar b \to H^+$ production with $H^+ \to t\bar b$, $W^+ h$ was studied in~\cite{He:1998ie,Diaz-Cruz:2001igs, Diaz-Cruz:2009ysj}}, and also with $H^+ \to \tau^+\nu$ in~\cite{Diaz-Cruz:2001igs, Diaz-Cruz:2009ysj, Slabospitsky:2002gw,Hernandez-Sanchez:2012vxa}.
We study resonant $H^+$ production followed by $H^+ \to W^+ H$ decay, together with $H \to t\bar c$ (see Fig.~\ref{diag}). Assuming the leptonic $t \to \ell^+\nu b$ ($\ell = e, \mu$) decay mode, we perform a signal-to-background analysis at the 14~TeV LHC to show that LHC Run 2 data at hand might already be sufficient to discover the $H^+$. 

\vskip0.1cm
%%%%%%%%%%%%%%%%%%%%%%%%%%%%%%%%%%%%%%%%%%%%%%%%
{\it G2HDM.---}
%%%%%%%%%%%%%%%%%%%%%%%%%%%%%%%%%%%%%%%%%%%%%%%%
The G2HDM has two weak scalar doublets with the same quantum numbers.
In the Higgs basis where only one doublet breaks the symmetry, the most general {\it CP}-conserving Higgs potential is~\cite{Davidson:2005cw,Hou:2017hiw}
\begin{align}
 & V(\Phi,\Phi') = \mu_{11}^2|\Phi|^2 + \mu_{22}^2|\Phi'|^2
    - (\mu_{12}^2\Phi^\dagger\Phi' + \rm{H.c.}) \\
 & \qquad + \frac{\eta_1}{2}|\Phi|^4 + \frac{\eta_2}{2}|\Phi'|^4
   + \eta_3|\Phi|^2|\Phi'|^2  + \eta_4 |\Phi^\dagger\Phi'|^2 \nn \\
 & \qquad + \left[\frac{\eta_5}{2}(\Phi^\dagger\Phi')^2
   + \left(\eta_6 |\Phi|^2 + \eta_7|\Phi'|^2\right) \Phi^\dagger\Phi' + \rm{H.c.}\right],\nn
%\label{pot}
\end{align}
where the quartic $\eta_i$ couplings are {\it real}, and $\Phi$ breaks the EW symmetry spontaneously through a nonzero vacuum expectation value---namely $\mu_{11}^2 = - \frac{1}{2}\eta_1 v^2$---while $\left\langle \Phi'\right\rangle = 0$, hence $\mu_{22}^2 > 0$. 
A second minimization condition eliminates $\mu_{12}^2$ as a parameter. One can diagonalize the $h$, $H$ mass-squared matrix by a mixing angle $\gamma$ ($\equiv \beta-\alpha$ in 
2HDM-II convention), which satisfies $s_{\gamma}c_{\gamma} = \eta_6 v^2/(m_H^2-m_h^2)$ \cite{Davidson:2005cw,Hou:2017hiw}, with $c_\gamma \equiv \cos\gamma$ ($s_\gamma \equiv \sin\gamma$).
In the small-$c_\gamma$ limit, known as ``alignment,'' $h$ resembles the SM Higgs boson. %indicating that the $h$--$H$ mixing angle $c_\gamma$ is rather small. 

%The Higgs masses can be written in terms of the potential parameters in Eq.~(1),
%
%\begin{align}
% & m_{H^+}^2 = \mu_{22}^2 +  \frac{1}{2}\eta_3 v^2,\\	
% & m_{A}^2 =  m^2_{H^+} + \frac{1}{2}(\eta_4 - \eta_5) v^2,\\
% & m_{H,h}^2 = \frac{1}{2}\bigg[m_A^2 + (\eta_1 + \eta_5) v^2\nn\\
% & \quad\quad \quad\quad \pm \sqrt{\left(m_A^2+ (\eta_5 - \eta_1) v^2\right)^2
%   + 4 \eta_6^2 v^4}\bigg].	
%\end{align}
%

%
\begin{figure*}[t!]
	\centering
	\includegraphics[width=.9 \textwidth]{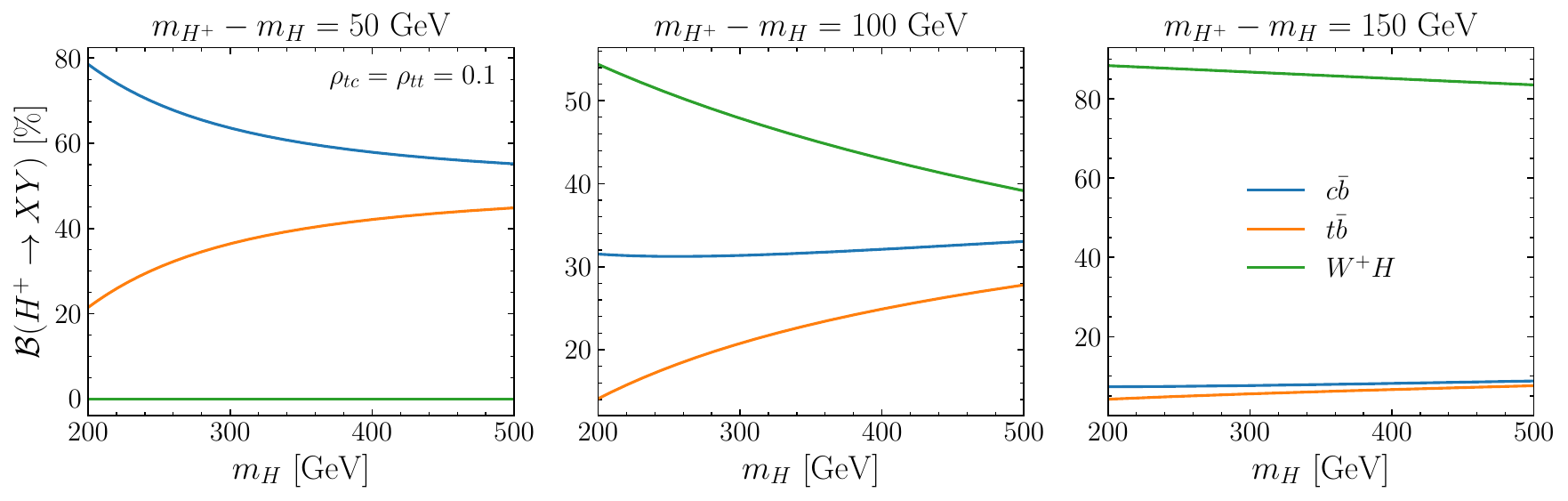}
	\vskip-0.1cm
	\caption{Branching ratios of $H^+$ as a function of $m_H$ for $m_{H^\pm} - m_H =$ 50 (left), 100 (middle), 150 GeV (right).}
	\label{Hdecay}
\end{figure*} 

The Yukawa Lagrangian is~\cite{Davidson:2005cw,Hou:2019mve}
\begin{align}
 \mathcal{L}_Y = %\supset
 & - \frac{1}{\sqrt{2}} \sum_{f = u, d, \ell}
       \bar f_{i} \bigg[\big(\lambda^f_{ij} c_\gamma + \rho^f_{ij} s_\gamma\big)H \nn \\
 & -i\,{\rm sgn}\,Q_f\, \rho^f_{ij} A - \big(\lambda^f_{ij} s_\gamma
       - \rho^f_{ij} c_\gamma\big) h  \bigg]  R f_{j}\nn \\
 & - \bar{u}_i\big[(V\rho^d)_{ij} R - (\rho^{u\dagger}V)_{ij} L\big]d_j H^+ \nn \\
 & - \bar{\nu}_i\rho^\ell_{ij} R \ell_j H^+ +{\rm H.c.},
\label{Lyuk}
\end{align}
where $i,j = 1$--3 are generation indices, $\rm{sgn}\,Q_f = +1\;(-1)$ for $f=u$ ($f=d,\ell$), 
$L,R = (1\mp\gamma_5)/2$ and $V$ is the CKM matrix. The $3\times3$ matrices 
$\lambda^f_{ij} \equiv \delta_{ij}\sqrt{2}m_i^f/v$, with $v \simeq 246~{\rm GeV}$, are diagonal and real, while $\rho^f_{ij}$ are nondiagonal and in general complex. 
For our collider study, we simplify and take $\rho^f_{ij}$ matrices as real.

The leading $\bar c bH^+$ coupling is $\rho_{tc}V_{tb}$ [see Eq.~(2)], where CKM suppression of 2HDM-II is absent~\cite{Ghosh:2019exx,He:1998ie}. In this paper, we focus on $c\bar b \to H^+ \to W^+ H$ (see Fig.~\ref{diag}), with $W^+ \to \ell^+ \nu$ and $H \to t\bar c \to  \ell^+ \nu b \bar c$, and we study the resulting same-sign dilepton signal at the 14~TeV LHC.

\vskip0.1cm
%%%%%%%%%%%%%%%%%%%%%%%%%%%%%%%%%%%%%%%%%%%%%%%%
{\it Parameter space.---}
%%%%%%%%%%%%%%%%%%%%%%%%%%%%%%%%%%%%%%%%%%%%%%%%
The parameters in Eq.~(1) are required to satisfy vacuum stability, tree-level unitarity and perturbativity conditions. To generate G2HDM parameters, we first express the quartic couplings $\eta_1$, $\eta_{3-6}$ in terms of $m^2_{h,H,A,H^+},~\mu^2_{22},~\gamma$, and $v$ \cite{Davidson:2005cw}, as these couplings enter the Higgs masses, while $\eta_2$ and $\eta_7$ do not. Since $H^+$ coupling to fermions is $c_\gamma$-independent [see Eq.~(2)], we set $c_\gamma = 0$ ($s_\gamma = 1$) and fix $m_h = 125$~GeV {to simplify}, hence $\eta_1 = m^2_h/v^2 \cong 0.258$ and $\eta_6 = 0$. Note that the bosonic $H^+$ couplings such as $hW^+H^-$ ($HW^+H^-$) are $c_\gamma$ ($s_\gamma$) suppressed (enhanced). The coupling $AW^+H^-$ is $\gamma$ independent. We then uniformly scan $m_H$ from 200 to {500}~GeV, assuming custodial symmetry ($m_{A} = m_{H^+}$) and $m_{H^+} = m_H + m^\prime$ ($m^\prime = 50, 100, 150$~GeV) to allow for $H^+ \to W^+ H$ decay, and we randomly scan remaining parameters within the following ranges: $\left|\eta_{2-5,7}\right| < 5$ (vacuum stability requires $\eta_2 > 0$) and $\mu^2_{22} \in [0, 10^6]$ GeV$^2$ using the \texttt{2HDMC-1.8.0} code~\cite{Eriksson:2009ws}, which is also used to impose theoretical restrictions. We note that constraints from electroweak precision $S,T,$ and $U$ observables, which require $m_{H^+} \simeq m_A$ (custodial, or $m_{H^+} \simeq m_H$ for twisted custodial~\cite{Gerard:2007kn}), are automatically satisfied.

As a scan result, we give $H^+$ decays to different final states---in particular $c\bar b$, $t\bar b$, and $W^+ H$ pairs. {For simplicity, we set all $\rho_{ij} = 0$ except $\rho_{tc}$ and $\rho_{tt}$. For $\rho_{tc} = \rho_{tt} = 0.1$, we plot in Fig.~\ref{Hdecay} the branching ratios of $H^+ \to c\bar b$, $ t\bar b$, and $W^+ H$} as functions of $m_H$, for $m_{H^+} - m_H = 50$~(left), 100~(middle) and 150~GeV~(right). 
(Note theoretical restrictions and the $\left|\eta_{3-5}\right| < 5$ requirement are not enforced for better illustration.)
We see that when there is not enough phase space ({$m_{H^+} - m_H \leq 50$ GeV}), $H^+ \to W^+ H$ is negligible and is hence set to 0, and $c\bar b$ decay (with $t\bar b$ subdominant) dominates $H^+$ decays. However, if $m_{H^+} - m_H > 100$~ GeV, $H^+ \to W^+ H$ would dominate with $W$ on-shell (middle and right of Fig.~2). In contrast, for $\rho_{tc} = 0.3, \rho_{tt} = 0.5$, ${\cal B}(H^+ \to t\bar b) > 0.5$ for $m_{H^+} - m_H \geq 100$~GeV, while $H^+ \to W^+ H$ could become important only for $m_{H^+} - m_H \geq 200$~GeV.
%\tcb{Additionally, for $\rho_{tc}, \rho_{tt}\,\sim\,0.5$ (as used in Refs.~\cite{Ghosh:2019exx,Hou:2024bzh}), ${\cal B}(H^+ \to t\bar b) > 0.5$ for $m_{H^+} - m_H \geq 100$~GeV}.

\vskip0.1cm
%%%%%%%%%%%%%%%%%%%%%%%%%%%%%%%%%%%%%%%%%%%%%%%%
{\it Constraints on $\rho_{tc}$.---}
%%%%%%%%%%%%%%%%%%%%%%%%%%%%%%%%%%%%%%%%%%%%%%%%
The top-charm coupling $\rho_{tc}$ is constrained by flavor physics and also direct searches.  
Flavor observables such as $B_s$-$\bar{B}_s$ mixing and $b \to s\gamma$ do not severely
constrain $\rho_{tc}$ due to the small $m_c$. These observables receive $\rho_{tc}$ contributions through charm-$H^+$ loops~\cite{Crivellin:2013wna}. We reinterpret the limits from Ref.~\cite{Crivellin:2013wna} and find that $\left|\rho_{tc}\right| \gtrsim 1.3~(1.7)$ is excluded by $B_s$-$\bar{B}_s$ for $m_{H^+} = 300~(500)$~GeV. We refer to Refs.~\cite{Crivellin:2013wna,Altunkaynak:2015twa} for more discussion on constraining $\rho_{tc}$ and $\rho_{tt}$.

The extra top Yukawa coupling $\rho_{tc}$ receives constraints from $t \to ch$ searches at the LHC. For $c_\gamma \neq 0$, limits are significant, where both ATLAS \cite{ATLAS:2024mih} and CMS \cite{CMS:2021hug,CMS:2024ubt} set 95\% CL limits using full Run 2 data. We find that $\left|\rho_{tc}\right| \gtrsim 0.5$ is excluded at a 95\% CL for $c_\gamma = 0.1$. The limit diminishes for $c_\gamma < 0.1$ and vanishes for $c_\gamma = 0$, which we assume.  

LHC data further constrain $\rho_{tc}$. It was found~\cite{Ghosh:2019exx} (and references therein) that the control region {for the $t\bar t W$ (CRW) background in the CMS $4t$ search}~\cite{CMS:2019rvj}, defined by a same-sign dilepton (electron or muon), transverse missing energy and no more than five jets, with two $b$-tagged, gives the most severe constraint on $\rho_{tc}$.
It was shown~\cite{Ghosh:2019exx} that finite $\rho_{tc}$ induces the $cg \to tH/A \to tt\bar c$ process and can feed {CRW of Ref.~\cite{CMS:2019rvj}} if both top quarks decay semileptonically, yielding a final state with a same sign dilepton and $2b$ plus extra jet {that is} almost identical to CRW. It was found that $\left|\rho_{tc}\right| \gtrsim 0.4$ is excluded for $m_H = 272$ ($m_A = 372$) GeV. The signal region SR12 of the CMS $4t$ study \cite{CMS:2019rvj} constrains $\rho_{tc}$ as well. We refer to Ref. \cite{Ghosh:2019exx} for more discussion.

Recent ATLAS searches \cite{ATLAS:2023tlp} for heavy Higgs bosons in multilepton plus $b$-jet final states, and CMS searches \cite{CMS:2023xpx} for exotic neutral Higgs bosons via $pp \to tH/A \to tt\bar c$ and $pp \to tH/A \to tt\bar u$ set constraints on $\rho_{tc}$ (and $\rho_{tt}$, $\rho_{tu}$, as well). Our chosen benchmark points (BPs) (see Table \ref{table:BPs}) satisfy all constraints stemming from {these}~LHC~searches. 

\vskip0.1cm
%%%%%%%%%%%%%%%%%%%%%%%%%%%%%%%%%%%%%%%%%%%%%%%%
%\noindent{\it Collider probes.---}
{\it Collider study.---}
%%%%%%%%%%%%%%%%%%%%%%%%%%%%%%%%%%%%%%%%%%%%%%%%
We study the same sign-dilepton signal at the LHC, {with at least two jets, at least one $b$-tagged,} $c\bar b \to H^+ \to W^+ H \to \ell^+ \nu t(\to \ell^+\nu b)\bar c + \rm{c.c.}$, in comparison to Refs.~\cite{Ghosh:2019exx,Hou:2024bzh}. We select two BPs with $m_{H} = 200,\,300$~GeV and $m_{H^+} = m_A = 300,\,500$~GeV ({see Table}~\ref{table:BPs}), {with $H^+\to W^+ H$ and $H\to t\bar{c} +\bar t c$}. The BPs satisfy all constraints mentioned. 

\begin{table}[t!]
	\centering
	{\small  \begin{tabular}{l c c c c  c  c c  c c} %| c| c
			\hline\hline
			% &&&&&&&&& \\ %&&
			BP & %$\eta_1$ &
			$\eta_2$ & $\eta_3$ & $\eta_4$ & $\eta_5$ %& $\, \eta_6 \,$
			& $\eta_7$ & $m_{H}$ & $m_A$ & $m_{H^+}$ & ${\mu_{22}^2/v^2}$  \\
			\hline
			1 & %0.26 &
			1.40 & 2.00 & {$-0.82$} & {$-0.82$} & %0 &
			{$-0.55$} & 200 & 300 & 300 & 0.49 \\
			2 & %0.26 &
			2.88 & 4.75 & {$-2.64$} & {$-2.64$} & %0 &
			\,\,\,~0.16 & 300 & 500 & 500 & 1.75 \\
			%   &&&&&&&&& \\ %&&
			\hline\hline
			%\hline
	\end{tabular}}
	\caption{G2HDM parameters for selected BPs.
		All masses in GeV, with $\eta_6 = 0$ and $m_h = 125$~GeV. For BP1, $\rho_{tc} = \rho_{tt} = 0.1$, while for BP2, $\rho_{tc} = 0.3,\,\rho_{tt} = 0.5$.}
	\label{table:BPs}
\end{table} 

There are two main background sources for our signal. 
The first category is irreducible,\footnote{We define an irreducible background as any process that produces one same-charge dilepton and a minimum of two jets, at least one of which is $b$-tagged.} mainly from $t\bar tV$ (where $V$ refers to either a $W$ or $Z$ boson) and $tZj$, with $t\bar th$ and $4t$ subdominant. Other subdominant backgrounds such as $t\bar tWW$ and $tt\bar{t}$ are neglected in our analysis. The second category is a reducible background, where a fake same-sign dilepton is produced in the detector through events containing electrons with misidentified charge (Q-flip) with $10^{-3}$ {probability}~\cite{ATLAS:2018alq, ATLAS:2016kjm, Alvarez:2016nrz, CMS:2019rvj}, and those including fake leptons (Fake) with $10^{-4}$ rate~\cite{ATLAS:2018alq, ATLAS:2016kjm, Alvarez:2016nrz, CMS:2019rvj}, mainly from $t\bar t +\rm{jets}$~\cite{ATLAS:2018alq, ATLAS:2016kjm, Alvarez:2016nrz, CMS:2019rvj}. We note that our mis-identification probabilities and fake rates are conservative. Relevant $tW$, $WZ$ and $ZZ$ backgrounds are classified as reducible. 
 
We generate signal and background events at leading order (LO) using \texttt{MadGraph5\_aMC@NLO}~\cite{Alwall:2014hca} with a default NN23LO1 \cite{Ball:2013hta} parton distribution function (PDF) set, 
interfaced {with} \texttt{Pythia-8.2}~\cite{Sjostrand:2014zea} for showering and hadronization. We then use \texttt{Delphes-3.5.0} \cite{deFavereau:2013fsa} with a default ATLAS card and anti-kt algorithm \cite{Cacciari:2008gp} with radius parameter $R = 0.5$ for detector simulation.
We consider one additional parton for ${t\bar t}W$, ${t\bar t}Z$, $tW$, $WZ$, and $ZZ$ background events, and two additional partons for $t\bar t+\rm{jets}$ %and $Z+\rm{jets}$ 
using the MLM matching scheme~\cite{Alwall:2007fs}. The $tZj$, ${t\bar t}h$, and $4t$ backgrounds are estimated without extra partons. We rescale background cross sections using the $K$-factor method to account for higher-order QCD corrections. $K$ factors are 1.54 (1.50) \cite{LHCHiggsCrossSectionWorkingGroup:2016ypw} for $t\bar t W^-$ ($t\bar t W^+$), 
1.40 \cite{LHCHiggsCrossSectionWorkingGroup:2016ypw} for $t\bar tZ$, 1.44 \cite{Alwall:2014hca} for $tZj$, 1.26 \cite{LHCHiggsCrossSectionWorkingGroup:2016ypw} for $t\bar th$
and 2.04 \cite{Alwall:2014hca} for $4t$. The $tW$, $W^-Z$ ($W^+ Z$), $ZZ$ and $t\bar t+\rm{jets}$ %and $Z+\rm{jets}$
backgrounds are rescaled by factors of {1.35}~\cite{Kidonakis:2010ux}, 
{1.30 (1.26)}~\cite{Campanario:2010hp}, {1.72}~\cite{Cascioli:2014yka} and 
{1.84}~\cite{twiki}, respectively. Signal cross sections are kept at LO. 
Note that both signal and background events are generated at 14 TeV collision energy. 

Signal events are generated up to one merged jet using the MLM {scheme~\cite{Alwall:2007fs}; thus, subdominant $cg \to bH^+$, $c\bar b \to gH^+$, and $\bar bg \to \bar cH^+$ effects are included.} These contributions are promising, as they have an extra high-$p_T$ ($b$-)jet that could help suppress the background~\cite{Ghosh:2019exx,Hou:2024bzh}. 

To minimize backgrounds, we take a cut-based approach.
Reconstructed objects are subject to the following selections: a minimum of two jets, of which at least one is $b$-tagged, with $p_T > 20$ GeV and $\left|\eta\right|<2.5$, exactly two leptons (electron or muon) with the same charge (events with more than two leptons are vetoed), with a leading (subleading) lepton satisfying $p_T > 25$ ($20$) GeV with $\left|\eta\right|<2.5$, a separation of $\Delta R > 0.4$ between the two same-charge leptons ($\Delta R_{\ell\ell}$), and between any lepton and any jet ($\Delta R_{\ell j}$), a missing energy $E^{\rm{miss}}_{T}$ larger than 35 GeV, and scalar $p_T$ sum of all jets and the two same-sign leptons ($H_T$) less than 400 GeV.

\begin{table}[t]
	\centering
	\setlength{\tabcolsep}{10pt}
	{\small \begin{tabular}{l c} %| c| c 
			\hline\hline
			% &&&&&&&&& \\ %&&
			Background & Cross section  \\
			\hline
			$tW$ & 1.61 \\			
			$t\bar tW$ & 1.09 \\
			$WZ$ & 0.54 \\			
			$tZj$ & 0.40 \\
			$t\bar tZ$ & 0.10 \\		
			%$tW$ &  \\							
			$tth$ & 0.05 \\
			$ZZ$ & 0.02 \\			
			$4t$ & 0.0004 \\ 
			Q-flip & 0.0018 \\ 
			Fake & 0.0002 \\								
			\hline\hline
			%\hline
	\end{tabular}}
	\caption{Background cross sections (fb) after selection cuts.}
	\label{table:xsBKG}
\end{table} 
\begin{table}[b]
	\centering
	\setlength{\tabcolsep}{10pt}
	{\small \begin{tabular}{l c c} %| c| c 
			\hline\hline
			% &&&&&&&&& \\ %&&
			BP & Signal & $\mathcal{Z}$ at 300 fb$^{-1}$\\
			\hline
			
			1 & 3.72 & 7.4 \\
			2 & 4.62 & 8.8\\	
			\hline\hline
			%\hline
	\end{tabular}}
	\caption{Signal cross sections (fb) and significance ($\epsilon = 10\%$) for BPs after selection cuts.}
	\label{table:xssignal}
\end{table} 
\begin{figure*}[t]
	\centering
	\includegraphics[width=.9 \textwidth]{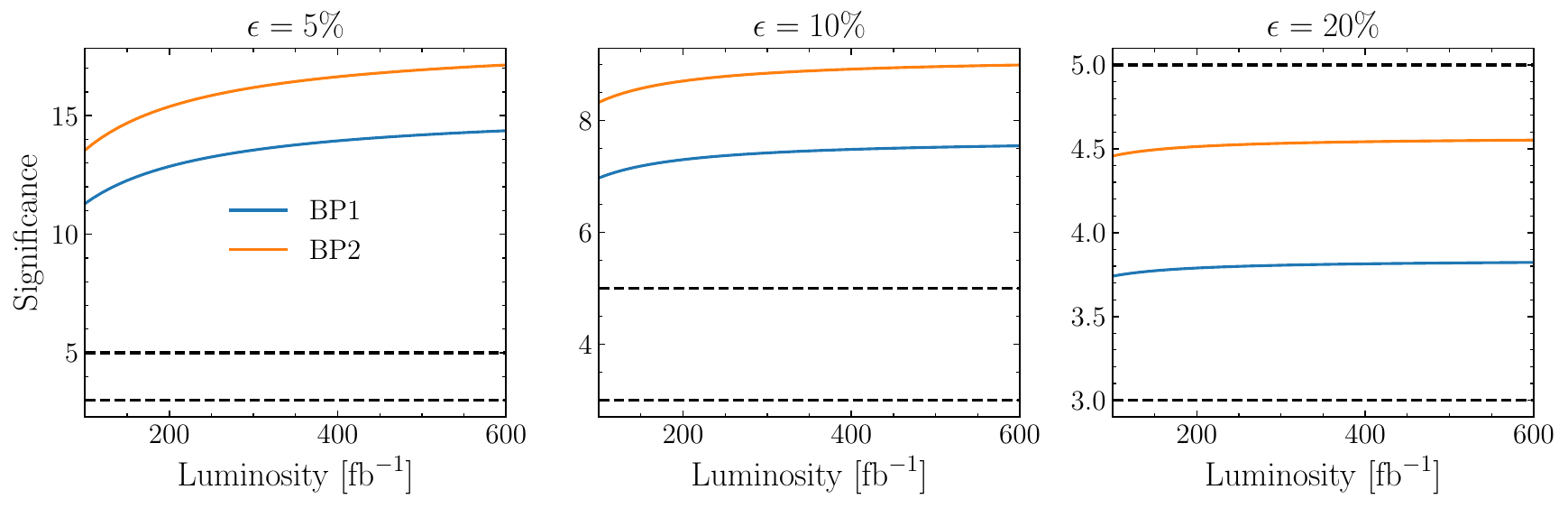}
	\vskip-0.1cm
	\caption{Significance as a function of luminosity for $\epsilon = 5\%$ (left),
              $\epsilon = 10\%$ (middle) and $\epsilon = 20\%$ (right).}
	\label{fig:significance}
\end{figure*}  

Background cross sections after selection cuts are given in Table~\ref{table:xsBKG}.
Signal cross sections for {the} BPs are shown in {Table}~\ref{table:xssignal} along with the significance calculated using the definition from {Eq.}~(1.4) of {Ref.}~\cite{Kumar:2015tna} (see also Ref.~\cite{Cowan:2010js}), which is analogous to but more accurate an estimate than $\mathcal{Z} \approx S/\sqrt{S+B+(\epsilon B)^2}$, where $S$ ($B$) is the number of signal (background) events, and $\epsilon$ refers to the systematic uncertainty in the background estimation. Assuming $\epsilon = 10\%$, 
the significance for $c\bar{b} \to H^+ \to W^+ H$ is $\sim$7.2$\sigma$ and 7.4$\sigma$ ($\sim$8.5$\sigma$, 8.8$\sigma$) for BP1 (BP2) at 140 and 300~fb$^{-1}$.
Thus, the full Run 2 data may be sufficient for discovery already.
Significances as a function of luminosity for systematic errors of $5\%$, $10\%$ and $20\%$ are plotted in Fig.~\ref{fig:significance}.

\vskip0.1cm
%%%%%%%%%%%%%%%%%%%%%%%%%%%%%%%%%%%%%%%%%%%%%%%%
{\it Discussion and Summary.---}
%%%%%%%%%%%%%%%%%%%%%%%%%%%%%%%%%%%%%%%%%%%%%%%%
There is a good chance that $H^+$ bosons can be discovered at the LHC with the Run~2 data already at hand. If so, it would not only shed light on the G2HDM, but also the mechanism behind BAU.   

We started with the conservative $\rho_{tc},\,\rho_{tt} \sim 0.1$~(BP1), rooted in our wish to keep $H^+ \to W^+ H$ dominant for $m_{H^+} - m_H \simeq 100$~GeV (see Fig.~\ref{Hdecay}),
and stressed that ${\rm Im}\,\rho_{tt} \cong 0.1$ is still quite robust for EWBG~\cite{Fuyuto:2017ewj}. Large $\rho_{tc},\,\rho_{tt}$ would enhance the production cross section but reduce ${\cal B}(H^+ \to W^+ H)$, {\it unless} a large~$m_{H^+}$-$m_H$~splitting is imposed.
For $\rho_{tc},\,\rho_{tt} \sim 0.4$ (BP2), $H^+ \to W^+ H$ would be larger than $t\bar b$ and $c\bar b$ decays if $m_{H^+} - m_H > 200$~GeV. With $m_H = 300$~GeV and $m_{H^+} = 500$~GeV, ${\cal B}(H^+ \to W^+ H)$, ${\cal B}(H^+ \to t\bar b)$, and ${\cal B}(H^+ \to c\bar b) \cong$ 49\%, 35\%, and 16\%, respectively.

For $\rho_{tc},\,\rho_{tt} \sim 0.5$ and $m_{H^+} \sim m_A \sim m_H$~\cite{Ghosh:2019exx,Hou:2024bzh}, $H^+ \to t\bar b$ {would be dominant}. However, the $c\bar b \to H^+ \to t\bar b$ signal would suffer from large $t\bar t$ and single-top backgrounds, and $c\bar b \to H^+ \to c\bar b$ would suffer from overwhelming QCD multijets backgrounds.
Therefore, $cg \to bH^+ \to bt\bar b$~\cite{Ghosh:2019exx} has been suggested
to search for $H^+$ bosons. The $\bar b g \to \bar cH^+ \to \bar ct\bar b$ 
signal~\cite{Hou:2024bzh} might also be useful, as it has a high-$p_T$ jet that could help suppress the background. But if $\rho_{tc},\,\rho_{tt}$ are small enough (e.g., {$\sim 0.1$}), these signals would be buried in the QCD background. Our proposed signal, $c\bar b \to H^+ \to W^+ H \to \ell^+ \nu t\bar c$, with its same-sign dilepton signature, could then be the most promising avenue for discovering $H^+$ at the LHC.

Turning on $\rho_{tu}$ would induce the $u\bar b \to H^+ \to W^+ H \to \ell^+ \nu t\bar c$ process without CKM suppression. However, $\rho_{tu}$ is severely constrained by flavor physics. It was found that $D$-$\bar{D}$ mixing and $B \to \tau\nu$ constrain $\left|\rho_{tc}\rho^*_{tu}\right| \lesssim 0.02$~\cite{Crivellin:2013wna} and $2.7\times 10^{-3} \lesssim \left|\rho_{tu}\right| \lesssim 2.0 \times 10^{-2}$~\cite{Crivellin:2013wna}, respectively, for $m_{H^+} = 500$ GeV. LHC data~\cite{ATLAS:2023tlp,CMS:2023xpx} put severe constraints on $\rho_{tu}$ as well. Taking $\rho_{tu} = 0.01$ ($\rho_{tc} = \rho_{tt} = 0.1$), the significance is $\sim$7.6$\sigma$ for BP1 at 300 fb$^{-1}$, slightly enhanced over $\rho_{tu} = 0$~(see Table \ref{table:xssignal}).
Note that the presence of $\rho_{tu}$ would also induce $H^+ \to u\bar b$ and $H \to t\bar u$ decays, but these are practically negligible.

Another important parameter worth mentioning is $s_\gamma$, which controls both $H^+ \to W^+ H$ and $H \to t\bar c$ decays. We set $s_\gamma = 1$ (alignment) in our analysis, but it is possible to work a little away from alignment. For example, setting $s_\gamma = 0.95$ ($c_\gamma \simeq 0.3$) would not significantly reduce the total cross sections, and hence significance. We rescale the signal cross sections and find $\mathcal{Z}\sim 6.3\sigma$ for BP1 at 300 fb$^{-1}$, assuming $\epsilon = 10\%$.

In this paper, we suggest searching for $H^+$ through the same-sign dilepton signal by  $c\bar b \to H^+ \to W^+(\to \ell^+\nu) H$, with $H \to t (\to \ell^+\nu b)\bar c$. We estimate the signal sensitivity by a signal-to-background analysis for $m_{H^+} =$~300--500~GeV.
Full Run~2 data may be sufficient for discovery already. One would not only discover the $H^+$ but also measure the top-charm coupling $\rho_{tc}$.
%which can drive electroweak baryogenesis.
We encourage ATLAS and CMS experiments to consider our suggested new avenue for the $H^+$ search.

%\newpage
\vskip0.2cm
\noindent{\bf Acknowledgments.--}
This work is supported by NSTC Grant No. 113-2639-M-002-006-ASP of Taiwan, and NTU Grants No. 113L86001 and No. 113L891801. We thank Abdesslam Arhrib for discussion.

%%%%%%%%%%%%%%%%%%%%%%%%%%%%%%%%%%%%%%%%%%%%%%%%%%%%%%%%%%%%%%%%%%%%%%%%%%%%%%%%%%%%%%%%%%

\end{document}